\documentclass[letter,twocolumn,epsf]{jpsj2} 
\title{Strong-Coupling Superconductivity of CeIrSi$_3$ \\with the Non-centrosymmetric Crystal Structure\thanks{This paper will be published in J. Phys. Soc. Jpn. on the August issue of 2007.}}

\author{Naoyuki \textsc{TATEIWA}$^{1}$\thanks{Email: tateiwa.naoyuki@jaea.go.jp}, Yoshinori \textsc{HAGA}$^{1}$, Tatsuma D. \textsc{MATSUDA}$^{1}$, Shugo \textsc{IKEDA}$^{1}$, Etsuji \textsc{YAMAMOTO}$^{1}$, Yusuke \textsc{OKUDA}$^{2}$, Yuichiro \textsc{MIYAUCHI}$^{2}$, Rikio \textsc{SETTAI}$^{2}$ and Yoshichika \textsc{\=ONUKI}$^{1,2}$}

\inst{$^{1}$Advanced Science Research Center, Japan Atomic Energy Agency, Tokai, Ibaraki 319-1195, Japan\\
$^{2}$Graduate School of Science, Osaka University, Toyonaka, Osaka 560-0043, Japan}

\abst{We studied the pressure-induced superconductor CeIrSi$_3$ with the non-centrosymmetric tetragonal structure under high pressure. The electrical resistivity and ac heat capacity were measured in the same run for the same sample. The critical pressure was determined to be $P_{\rm c}$ = 2.25 GPa , where the antiferromagnetic state disappears. The heat capacity $C_{\rm ac}$ shows both antiferromagnetic and superconducting transitions at pressures close to $P_{\rm c}$. On the other hand, the superconducting region is extended to high pressures of up to about 3.5 GPa, with the maximum transition temperature $T_{\rm sc}$ = 1.6 K around $2.5-2.7$ GPa. At 2.58 GPa, a large heat capacity anomaly was observed at $T_{\rm sc}$ = 1.59 K. The jump of the heat capacity in the form of ${\Delta}{C_{\rm ac}}/C_{\rm ac}(T_{\rm sc})$ is 5.7 $\pm$ 0.1. This is the largest observed value among previously reported superconductors, indicating the strong-coupling superconductivity. The electronic specific heat coefficient at $T_{\rm sc}$ is, however, approximately unchanged as a function of pressure, even at $P_{\rm c}$. 
 }

\kword{CeIrSi$_3$, ac calorimetry, strong-coupling superconductivity}

\begin{document}
\maketitle

  Recently, the discovery of non-centrosymmetric superconductors such as CePt$_3$Si, UIr, CeRhSi$_3$, CeIrSi$_3$, and CeCoGe$_3$ has attracted considerable interest~\cite{bauer,akazawa,sugitani,kimura,settai}. In non-centrosymmetric superconductors, the superconducting pairing state is considered as an admixture of the spin singlet and triplet channels imposed by the presence of antisymmetric spin-orbit coupling based on the crystal structure without inversion symmetry~\cite{mineev}. Many theoretical and experimental studies have been carried out extensively in order to clarify this novel type of unconventional superconductivity~\cite{onuki}. 
 
   In this letter, we report our high pressure study on CeIrSi$_3$, which crystalizes in the BaNiSn$_3$-type tetragonal crystal structure ($I$4$mm$). The Ce atoms occupy the four corners and the body center of the tetragonal structure in a manner similar to the well-known tetragonal ThCr$_{2}$Si$_{2}$-type structure, but the arrangement of the Ir and Si atoms lacks inversion symmetry along the [001] direction ({\it c} axis). At ambient pressure, CeIrSi$_3$ shows an antiferromagnetic transition at a N\'{e}el temperature $T_{\rm N}$ = 5.0 K. The N\'{e}el temperature decreases monotonically with increasing pressure and disappears at a critical pressure $P_{\rm c}$ = 2.25 GPa, as shown in Fig. 1, where the critical pressure $P_{\rm c}$ (= 2.25 GPa) was determined from the present experiment shown later in combination with the previous experimental results~\cite{sugitani,okuda}. Superconductivity appears in a wide pressure region from 1.3 GPa to about 3.5 GPa, with a maximum superconducting transition temperature $T_{\rm sc}$ = 1.6 K around 2.5 GPa.  We performed the ac heat capacity and electrical resistivity measurements on CeIrSi$_3$ in order to study the superconductivity. 
       
        \begin{figure}[b]
\begin{center}

\includegraphics[width=7.5cm]{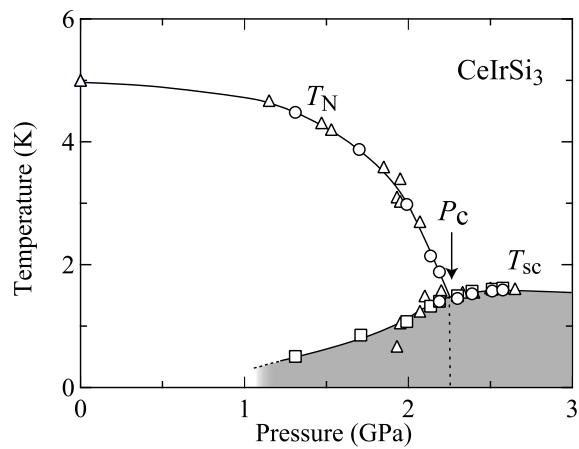}
 \end{center}
\caption{\label{fig:epsart} Pressure phase diagram of CeIrSi$_3$. $T_{\rm N}$ and $T_{\rm sc}$, which were determined by the previous resistivity measurements, are plotted by triangles~\cite{sugitani,okuda}. $T_{\rm sc}$ and $T_{\rm N}$ values obtained by the present resistivity and ac heat capacity measurements are plotted by squares and circles, respectively.}
\end{figure} 
  The single crystal of CeIrSi$_3$ was grown by the Czochralski method in a tetra-arc furnace. The details of the sample preparation are given in the recent paper\cite{okuda}. The residual resistivity ratio RRR (= ${\rho}_{\rm RT}/{\rho}_{\rm 0}$) is 120, where ${\rho}_{\rm RT}$ and ${\rho}_{\rm 0}$ are the resistivity at room temperature and the residual resistivity, respectively, indicating the high quality of the sample. The heat capacity under high pressures was measured by the ac calorimetry method~\cite{wilhelm,tateiwa}.  The sample was heated up using a heater, whose power is modulated at a frequency ${\omega}$. The amplitude of the temperature oscillation $T_{\rm ac}$ is related to the heat capacity $C_{\rm ac}$ of the sample: $ T_{\rm ac} = P_{0}/({\kappa}+i{\omega}C_{\rm ac})$. Here, $P_{0}$ is an average of the power. ${\kappa}$ is the thermal conductivity between the sample and the environment.  $T_{\rm ac}$ was measured with a AuFe/Au thermocouple (Au + 0.07 at\% Fe). The contribution from the thermocouple and Au wires to the heat capacity is very small ($\sim$ 0.1\%).  The resistivity measurement was also carried out for the same sample by the standard four-terminal method.  For the resistivity measurement, two additional Au wires were attached to the edges of the sample so as to pass the electrical current. The low-temperature measurement was carried out using a $^{3}$He refrigerator from 0.3 K to 10 K. We used a hybrid piston cylinder-type cell. Daphne oil (7373) was used as a pressure transmitting medium.

  \begin{figure}[t]
\begin{center}

\includegraphics[width=7.5cm]{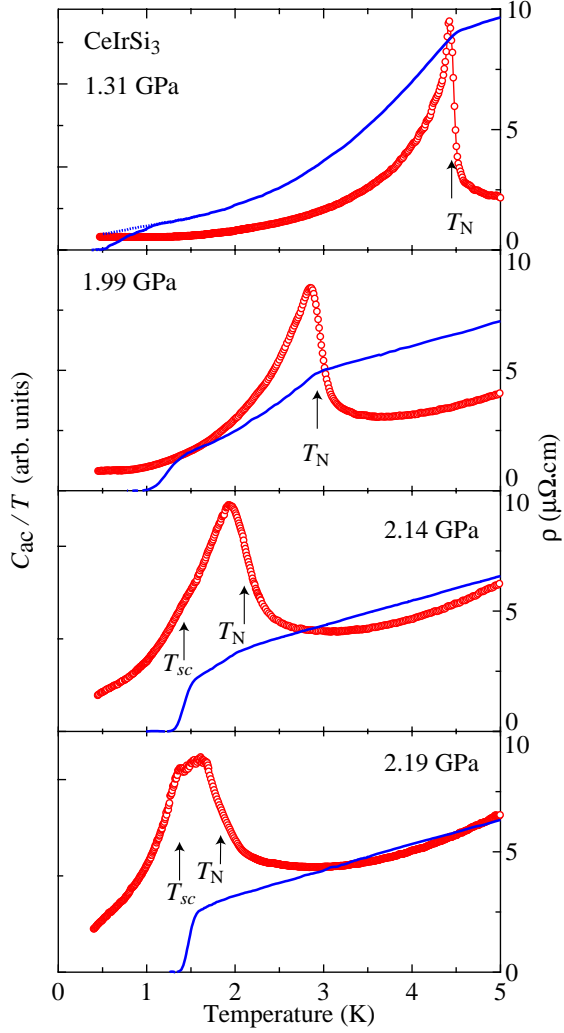}  
 \end{center}
\caption{\label{fig:epsart} (Color online) Temperature dependences of the ac heat capacity $C_{\rm ac}$ (circles, left side) and electrical resistivity $\rho$ (lines, right side) at 1.31, 1.99, 2.14, and 2.19 GPa in CeIrSi$_3$.}
\end{figure}


  Figure 2 shows the temperature dependences of the heat capacity $C_{\rm ac}$ and electrical resistivity $\rho$ below the critical pressure $P_{\rm c}$ = 2.25 GPa. At 1.31 GPa, $C_{\rm ac}$ shows a clear anomaly and $\rho$ shows a kink at the N\'{e}el temperature $T_{\rm N}$ = 4.48 K. The temperature dependence of $\rho$ at lower temperatures depends on the applied electrical current $j$. The data for $j$ = 1.0 and 0.3 A/cm$^2$ are shown by the dotted and solid lines, respectively. The resistivity reveals a superconducting transition at $T_{\rm sc}$ = 0.51 K for $j$ = 0.3 A/cm$^2$, while no transition for $j$ = 1.0 A/cm$^2$. There is, however,  no superconducting anomaly in $C_{\rm ac}$  at $T_{\rm sc}$.  At 1.99 GPa, $C_{\rm ac}$ and $\rho$ show a clear transition at $T_{\rm N}$ = 2.95 K, and only $\rho$ shows the superconducting transition at $T_{\rm sc}$ = 1.02 K. The evidence of the bulk superconductivity is not obtained at 1.31, 1.71 (data not shown), and 1.99 GPa. At 2.14 GPa, the antiferromagnetic transition is clearly indicated in both $\rho$ and $C_{\rm ac}$, while $\rho$ shows the superconducting transition at $T_{\rm sc}$ = 1.31 K, and $C_{\rm ac}$ shows a weak hump around $T_{\rm sc}$.  At 2.19 GPa, $C_{\rm ac}$ shows a broad anomaly with two peak structures, which correspond to the antiferromagnetic and superconducting transitions, respectively. The N\'{e}el temperature is determined as $T_{\rm N}$ = 1.88 K from the entropy balance. The peak of the heat capacity at the lower temperature side is close to the superconducting transition at $T_{\rm sc}$ = 1.40 K, where $\rho$ becomes zero.

    \begin{figure}[t]
\begin{center}

 \includegraphics[width=7.5cm]{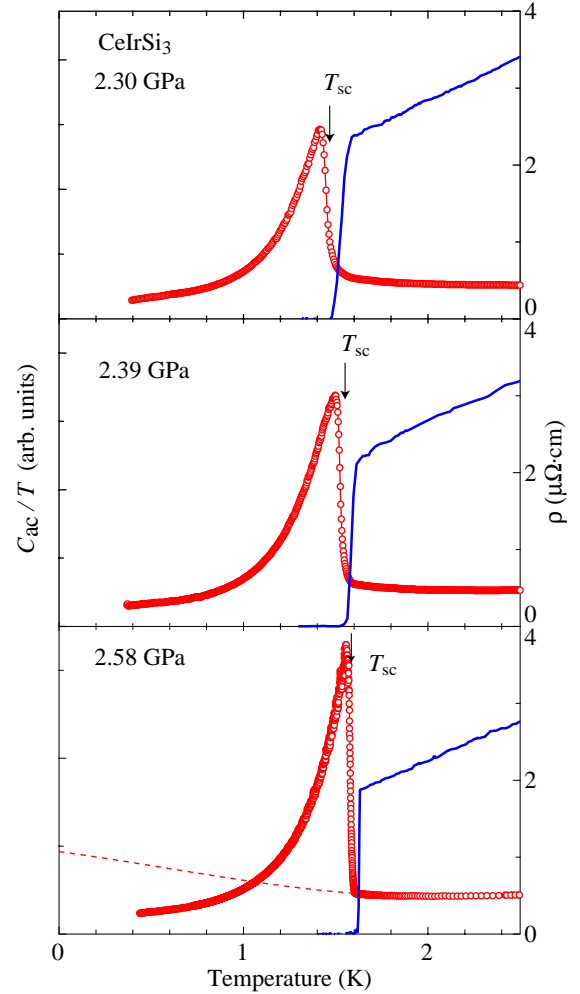}
 \end{center}
\caption{\label{fig:epsart} (Color online) Temperature dependences of the ac heat capacity $C_{\rm ac}$ (circles, left side) and electrical resistivity $\rho$ (lines, right side) at 2.30, 2.39, and 2.58 GPa in CeIrSi$_3$. The dotted line indicates the entropy balance below $T_{\rm sc}$ at 2.58 GPa.}
\end{figure} 
  
  At pressures higher than $P_{\rm c}$ = 2.25 GPa, only the superconducting transition is observed in both $C_{\rm ac}$ and $\rho$, as shown in Fig. 3. At 2.58 GPa, the values of $T_{\rm sc}$ obtained from the resistivity and ac heat capacity measurements are 1.62 and 1.59 K, respectively, which are plotted in Fig. 1 as squares and circles, respectively. The jump of the heat capacity in the form of ${\Delta}{C_{\rm ac}}/C_{\rm ac}(T_{\rm sc})$ is 3.4 $\pm$ 0.3 at 2.30 GPa, 4.40 $\pm$ 0.1 at 2.39 GPa, and 5.7 $\pm$ 0.1 at 2.58 GPa. Here, $\it{\Delta{C_{\rm ac}}}$ is the jump of the heat capacity at $T_{\rm sc}$, and $C_{\rm ac}(T_{\rm sc})$ is the value of $C_{\rm ac}$ just above $T_{\rm sc}$, namely, corresponding to ${\gamma}T_{\rm sc}$, where ${\gamma}$ is the electronic specific heat coefficient. The values of $\it{\Delta{C_{\rm ac}}/C_{\rm ac}(T_{\rm sc})}$ are extremely larger than the BCS value of 1.43. In particular, the value of 5.7 $\pm$ 0.1 at 2.58 GPa is the largest value among the values reported for studied superconductors. Here, we considered the entropy balance in the superconducting state of 2.58 GPa, as shown by the dotted line. The value of $C_{\rm ac}/T$ is enhanced with decreasing temperature. The value of $C_{\rm ac}/T$ at 0 K is roughly twice larger than that at $T_{\rm sc}$ = 1.59 K. If the $C_{\rm ac}/T$ value at 0 K is used as the $\gamma$ value,  ${\Delta}{C_{\rm ac}}/({\gamma}T_{\rm sc})$ is around 2.8 $\pm$ 0.3. This is still larger than the BCS value. The precise pressure dependence of  ${\Delta}{C_{\rm ac}}/C_{\rm ac}(T_{\rm sc})$ is shown in Fig. 4(a). The strong-coupling superconductivity is thus realized at around 2.5 GPa in CeIrSi$_3$.  The increment of ${\Delta}{C_{\rm ac}}/C_{\rm ac}(T_{\rm sc})$ suggests that the superconducting coupling parameter increases with increasing pressure~\cite{scalapino,carbotte,bulaevskii}.

 \begin{figure}[t]
\begin{center}
 \includegraphics[width=7.5cm]{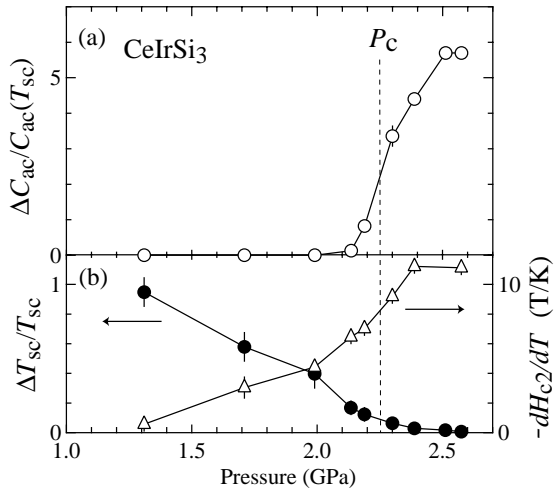}
 \end{center}
\caption{\label{fig:epsart} Pressure dependences of (a) ${\Delta}{C_{\rm ac}}/C_{\rm ac}(T_{\rm sc})$, and (b) the width of the superconducting transition in the resistivity ${\it{\Delta}}T_{\rm sc}$/$T_{\rm sc}$ (closed circles, left side) and the slope of the upper critical field $-dH_{\rm c2}/dT$ at $T_{\rm sc}$ (triangles, right side).}
\end{figure} 

  As shown in Fig. 4(b), the superconducting transition width ${\it{\Delta}}T_{\rm sc}$/$T_{\rm sc}$ in the resistivity decreases as a function of pressure and is close to zero above $P_{\rm c}$ = 2.25 GPa. Correspondingly, the slope of the upper critical field $H_{\rm c2}$ at $T_{\rm sc}$, $-dH_{\rm c2}/dT$, which is determined from the resistivity measurement in magnetic fields, becomes large: $-dH_{\rm c2}/dT$ = 11.2 T/K at 2.58 GPa for the field along the [110] direction, as shown in Fig. 4 (b). From these data shown in Fig. 4, we point out two possibilities for the co-existence of superconductivity and antiferromagnetism. One is that the superconductivity co-exists with the antiferromagnetism in a small pressure region close to $P_{\rm c}$. The other possibility is that both states do not coexist and the superconductivity exists inhomogeneously below $P_{\rm c}$. It is noted that the pressure dependence of ${\it{\Delta}}T_{\rm sc}$/$T_{\rm sc}$ as well as the gradual increase of ${\Delta}{C_{\rm ac}}/C_{\rm ac}(T_{\rm sc})$ around $P_{\rm c}$ can be interpreted as the increment of the superconducting volume fraction.  For further investigations on the co-existence of antiferromagnetism and superconductivity, microscopic experiments such as NMR are needed. 
 
  In the present ac heat capacity measurement, the absolute value of the heat capacity is not obtained, but the relative change of the heat capacity can be estimated~\cite{wilhelm,tateiwa}.  The value of $C_{\rm ac}/T$ just above $T_{\rm sc}$ is determined as 100 $\pm$ 20 mJ/K${^2}{\cdot}$mol at 2.58 GPa by comparison with the value of $C_{\rm ac}$ at ambient pressure. This $\gamma$ value indicates that the moderate heavy fermion superconductivity is realized in CeIrSi$_3$. Surprisingly, this value is approximately the same as $\gamma$ = 120 or 105 mJ/K${^2}{\cdot}$mol at ambient pressure~\cite{okuda,muro}.  Figure 5 (a) shows the pressure dependences of $C_{\rm ac}/T$ at low temperatures. The data in the paramagnetic state are shown, which are normalized by the value at 2.58 GPa and are shifted upwards by one, two, and three scales for 3.0, 4.0, and 5.0 K, respectively, as compared to the data for 2.0 K. There is no distinct change in the pressure dependence of $C_{\rm ac}/T$. This result indicates that the electronic specific heat coefficient $\gamma$ is almost pressure independent, even at  the antiferromagnetic critical pressure $P_{\rm c}$ = 2.25 GPa. In the weak coupling limit,  $-dH_{\rm c2}/dT$ at $T_{\rm sc}$ is proportional to the square of the effective mass of the conduction electrons, $m^{*2}$~\cite{flouquet}. The large value of  $-dH_{\rm c2}/dT$ = 11.2 T/K at 2.58 GPa is not explained by the existence of conduction electrons with the large effective mass because the $\gamma$ ($\propto$ $m^{*}$) value is approximately unchanged as a function of pressure.  The large value of $-dH_{\rm c2}/dT$ at $T_{\rm sc}$ may be related to the enhancement of the superconducting coupling parameter~\cite{scalapino,carbotte,bulaevskii}.

 \begin{figure}[b]
\begin{center}
\includegraphics[width=7.5cm]{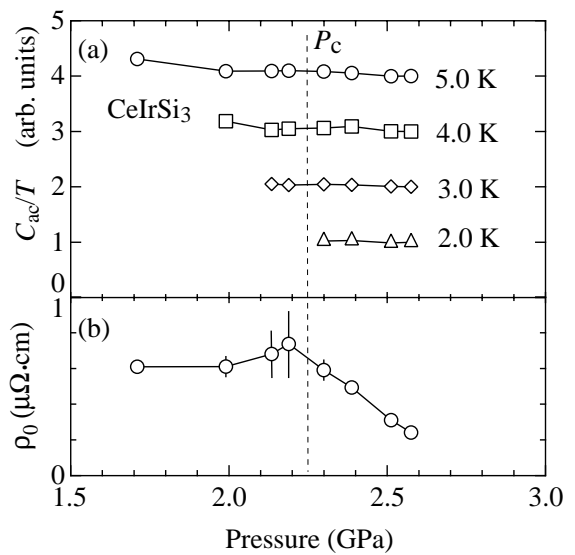}

 \end{center}
\caption{\label{fig:epsart} Pressure dependences of (a) $C_{\rm ac}/T$ at several temperatures and (b) the residual resistivity $\rho_{0}$. The  $C_{\rm ac}/T$ data for 3.0, 4.0, and 5.0 K are shifted up by one, two, and three scales, respectively, in (a) for clarity.}
\end{figure} 
   
  In the previous study, it was clarified that $\rho$ shows a $T$-linear temperature dependence at 2.5 GPa~\cite{sugitani}. In the present study, the $T$-linear temperature dependence of the resistivity is observed in a wide temperature region up to 25 K at pressures above the critical pressure $P_{\rm c}$.  Figure 5 (b) shows the pressure dependence of the residual resistivity $\rho_{0}$, which is almost constant below $P_{\rm c}$ = 2.25 GPa and starts to decrease considerably above $P_{\rm c}$. The value of $\rho_{0}$ at 2.58 GPa (0.24 ${\mu}{\Omega}{\cdot}$cm) is about 25 ${\%}$ of that at ambient pressure (0.96 ${\mu}{\Omega}{\cdot}$cm). 
  
  The relation between the antiferromagnetism and superconductivity is the most interesting issue to be discussed. From the present experimental results shown in Fig. 4, the superconductivity and antiferrromagnetism in CeIrSi$_3$ seem to be competing with each other. This is different from the case of the proto-type non-centrosymmetric superconductor CePt$_3$Si, where the superconductivity appears at $T_{\rm sc}$ = 0.75 K in the antiferromagnetic ordered state below $T_{\rm N}$ = 2.2 K~\cite{bauer}. Under high pressure, the value of ${\Delta}{C_{\rm ac}}/C_{\rm ac}(T_{\rm sc})$ starts to decrease and the superconducting transition width ${\it{\Delta}}T_{\rm sc}$/$T_{\rm sc}$ starts to increase above the critical pressure $P_{\rm c}$ $\sim$ 0.6 GPa~\cite{tateiwa,nicklas}. This suggests the cooperative relation between the two states in CePt$_3$Si. In both CeIrSi$_3$ and CePt$_3$Si, there is no anomalous behavior in the pressure dependence of the linear heat capacity coefficient $\gamma$ around the antiferromagnetic critical pressure. This is in strong contrast to other Ce-based pressure-induced superconductors such as CeIn$_3$ and CeRhIn$_5$~\cite{flouquet,hegger,shishido}. Superconductivity in CeIrSi$_3$ is different from superconductivity associated with the magnetic instability around the magnetic critical region.

  We compare the present experimental results with those of a pressure-induced superconductor CeRhIn$_5$ and a well-known strong coupling superconductor CeCoIn$_5$ (${\Delta}{C}/{\gamma}{T_{\rm sc}}$ = 4.5 )~\cite{hegger,petrovic,ikeda}. In CeRhIn$_5$,  the cyclotron effective mass obtained from the de Haas-van Alphen experiment and the residual resistivity $\rho_{0}$ indicate a divergent tendency around the critical pressure $P_{\rm c}$ = 2.35 GPa, where $T_{\rm sc}$ becomes maximum~\cite{muramatsu,shishido}. A marked change in the 4{\it f}-electron nature from localized to itinerant is realized at $P_{\rm c}$ under magnetic field. The value of ${\Delta}{C_{\rm ac}}/C_{\rm ac}(T_{\rm sc})$ shows a maximum value of 1.42 at $P_{\rm c}$~\cite{knebel}. In CeCoIn$_5$, ${\Delta}{C_{\rm ac}}/C_{\rm ac}(T_{\rm sc})$ decreases with increasing pressure from 4.5 at 0 GPa to 1.0 around 3 GPa~\cite{knebel}. Correspondingly, the values of  $\gamma$ and $\rho_{0}$ decrease considerably~\cite{sparn,sidorov}. In these two superconductors, the jump of the heat capacity at $T_{\rm sc}$, ${\Delta}{C_{\rm ac}}/C_{\rm ac}(T_{\rm sc})$, correlates with the enhancements of the $\gamma$ and $\rho_{0}$ values. The superconductivity is related to the formation of the heavy fermion state.

  On the other hand, in CeIrSi$_3$, no divergent tendency is observed in $\gamma$ and $\rho_{0}$ at $P_{\rm c}$ = 2.25 GPa and around $2.5-2.7$ GPa, where ${\Delta}{C_{\rm ac}}/C_{\rm ac}(T_{\rm sc})$ shows a maximum value. Theoretically, the non-centrosymmetric superconductivity of CeIrSi$_3$ needs to be considered, especially on the basis of the present experimental result that CeIrSi$_3$ is a strong-coupling superconductor with a moderate value of $\gamma$ = 100 $\pm$ 20 mJ/K${^2}{\cdot}$mol. Also the highly anisotropic and extremely large $H_{\rm c2} (0)$: $H_{\rm c2} (0)$ $\cong$ 30 T for $H$ $\|$ [001] and 9.5 T for $H$ $\|$ [110] at 2.65 GPa needs to be considered~\cite{okuda}. The upper critical field for $H$ $\|$ [001] is not suppressed by spin polarization based on the Zeeman coupling and possesses an upturn curvature below 1 K~\cite{okuda}. The upturn curvature may be related to the strong-coupling effect of superconductivity, as discussed in UBe$_{13}$~\cite{thomas}. The upturn curvature for the orbital critical field is expected when the coupling parameter $\lambda$ is large~\cite{bulaevskii}. As discussed in ref. 8, the Pauli paramagnetic effect for $H$ $\|$ [001] is strongly reduced in CeIrSi$_3$. Thus, the orbital effect becomes dominant in $H_{\rm c2}(T)$ for $H$ $\|$ [001] under low and moderate magnetic fields at low temperatures. The strong-coupling effect on the orbital critical field may be reflected in the observed upturn curvature in CeIrSi$_3$. It is interesting to note that a similar upturn curvature of the upper critical field for $H$ $\|$ [001] is also observed in CeRhSi$_3$~\cite{kimura2}.

     The present experimental results are summarized as follows. The critical pressure is determined to be $P_{\rm c}$ = 2.25 GPa, where the antiferromagnetic ordering disappears. Bulk superconductivity is only realized above $P_{\rm c}$. The highest $T_{\rm sc}$ = 1.6 K and ${\Delta}{C_{\rm ac}}/C_{\rm ac}(T_{\rm sc})$ = 5.7 $\pm$ 0.1 values are obtained at pressures higher than $P_{\rm c}$, namely, around $2.5-2.7$ GPa, indicating that CeIrSi$_3$ is a strong-coupling superconductor. The $\gamma$ value of 100 $\pm$ 20 mJ/K$^2$$\cdot$mol at $T_{\rm sc}$ is approximately unchanged as a function of  pressure.    
       
 \section*{Acknowledgements}
 One of the authors (N.T.) thanks Dr. G. Knebel for sending the heat capacity data of CeRhIn$_5$. This work was financially supported by the Grant-in-Aid for Creative Scientific Research (15GS0213), Scientific Research of Priority Area and Scientific Research (A and C) from the Ministry of Education, Culture, Sports, Science and Technology (MEXT).

\end{document}